\documentclass[12pt]{article}

\begin{document}

\begin{center}
{\Large\bf General relativistic tidal work for Papapetrou,
Weinberg and Goldberg pseudotensors}
\end{center}

\begin{center}
Lau Loi So  
\end{center}

\begin{abstract}
In 1998 Thorne claimed that all pseudotensors give the same tidal
work as the Newtonian theory. In 1999, Purdue used the
Landau-Lifshitz pseudotensor to calculate the tidal heating and
the result matched with the Newtonian gravity. Soon after in 2001,
Favata employed the same method to examine the Einstein,
Bergmann-Thomson and M{\o}ller pseudotensors, all of them give the
same result as Purdue did. Inspired by the work of Purdue and
Favata, for the completeness, here we manipulate the tidal work
for Papapetrou, Weinberg and Goldberg pseudotensors. We obtained
the same tidal work as Purdue achieved. In addition, we emphasize
that a suitable gravitational energy-momentum pseudotensor
requires fulfill the inside matter condition and all of the
classical pseudotensors pass this test except M$\o$ller. Moreover,
we constructed a general pseudotesnor which is modified by 13
linear artificial higher order terms combination with Einstein
pseudotensor. We find that the result agrees with Thorne's
prediction, i.e., relativistic tidal work is pseudotensor
independent.
\end{abstract}


\section{Introduction}
The tidal heating or tidal work is a real physical phenomenon.
Tidal heating means the net work done by an external tidal field
on an isolated body~\cite{Purdue}. A typical example is the
Jupiter-Io system. The satellite Io receiving a huge gravitational
energy through the tidal field from Jupiter, i.e., intensive
volcanism activities are observed~\cite{Peale}. But how to
calculate the tidal heating rate?  Pseudotensor may be one of the
appropriate options. In 1998 Thorne~\cite{Thorne} claimed that all
pseudotensors give the same relativistic tidal heating as the
Newtonian theory. The present paper is trying to illustrate that
Thorne's prediction is valid. More precisely, we have verified
that the tidal work rate is indeed independent of one's choice how
to localize the gravitational energy for all pseudotensors.

The tidal work rate is $\dot{W}=-\frac{1}{2}\dot{I}_{ij}E^{ij}$,
where $W$ refers to the tidal work, the dot means differentiate
w.r.t. time t, $I_{ij}$ is the mass quadrupole moment of the
isolated body and $E_{ij}$ is the tidal field of the external
universe. Note that both $I_{ij}$ and $E_{ij}$ are time dependent,
symmetric and traceless. Here we emphasize  that the physical
meaning of $\dot{W}$ is the rate at which the external field does
work on the isolated planet and this is an energy dissipation
process which means time irreversible~\cite{Booth}. In contrast,
there exists a recoverable process
$\dot{E}_{\rm{}int}\sim\frac{d}{dt}(I_{ij}E^{ij})$ which is time
reversible, where $E_{int}$ is the energy interaction between the
isolated planet's quadrupolar deformation and the external tidal
field.  Purdue uses $E_{\rm{}int}=\frac{\gamma+2}{10}I_{ij}E^{ij}$
to interpret different choices of energy localization by tuning
the coefficient $\gamma$~\cite{Purdue}.

It is known that there are infinite number of pseudotensors,
however, for the classical pseudotensors are several. They are
Einstein~\cite{Einstein}, Landau-Lifshitz~\cite{LL},
Bergmann-Thomson~\cite{BT}, Goldberg~\cite{Goldberg},
Papapetrou~\cite{Papapetrou}, Weinberg~\cite{Weinberg} and
M{\o}ller~\cite{Moller}. In 1999, Purdue~\cite{Purdue} used the
Landau-Lifshitz pseudotensor to calculate the tidal heating and
the result matched with the Newtonian perspective. Soon after that
in 2001, Favata~\cite{Favata} employed the same method to examine
the Einstein, Bergmann-Thomson and M{\o}ller pseudotensors, all of
them give the same result as Purdue did. Inspired by the work of
Purdue and Favata, for the completeness, here we manipulate the
tidal work for the Papapetrou, Weinberg and Goldberg
pseudotensors, we find that all of them give the same desired
tidal work as Purdue achieved.

In addition, we elucidate that a suitable gravitational
energy-momentum pseudotensor requires satisfy the inside matter
condition~\cite{MTW}, i.e., see (\ref{9bMar2015}).  It is known
that all of the classical pseudotensors pass this test exclude
M$\o$ller~\cite{CQGSoNesterChen2009}. Moreover, we constructed a
general pseudotesnor expression which is modified by 13 linear
artificial higher order terms combination with Einstein
pseudotensor. We find that all the results satisfy Thorne's
prediction, i.e., relativistic tidal work rate is pseudotensor
independent.

\section{Technical background}
Here we used the same spacetime signature and the notation as
in~\cite{MTW}: we set the geometrical units $G=c=1$, where $G$ and
$c$ are the Newtonian constant and speed of light.  The Greek
letters denote the spacetime and Latin letters refer to spatial.
In principle, the classical
pseudotensor~\cite{CQGSoNesterChen2009} can be obtained from a
rearrangement of the Einstein field equation:
$G_{\alpha\beta}=\kappa\,T_{\alpha\beta}$, where constant
$\kappa=8\pi{}G/c^{4}$, $G_{\alpha\beta}$ and $T_{\alpha\beta}$
are the Einstein and stress tensors.  We define the gravitational
energy-momentum density pseudotensor in terms of a suitable
superpotential $U_{\alpha}{}^{[\mu\nu]}$:
\begin{eqnarray}
2\kappa\sqrt{-g}\,t_{\alpha}{}^{\mu}:=\partial_{\nu}U_{\alpha}{}^{[\mu\nu]}-2\sqrt{-g}\,G_{\alpha}{}^{\mu}.
\label{6aMar2015}
\end{eqnarray}
Alternatively, one can rewrite (\ref{6aMar2015}) as
$\partial_{\nu}U_{\alpha}{}^{[\mu\nu]}
=2\sqrt{-g}{}(G_{\alpha}{}^{\mu}+\kappa\,t_{\alpha}{}^{\mu})$.
Using the Einstein equation, the total energy-momentum density
complex can be defined as
\begin{eqnarray}
{\cal{T}}_{\alpha}{}^{\mu}:=\sqrt{-g}(T_{\alpha}{}^{\mu}+t_{\alpha}{}^{\mu})
=(2\kappa)^{-1}\partial_{\nu}U_{\alpha}{}^{[\mu\nu]}.
\end{eqnarray}
This total energy-momentum density is automatically conserved as
$\partial_{\mu}{\cal{T}}_{\alpha}{}^{\mu}\equiv{}0$ which can be
classified into two parts
\begin{eqnarray}
\partial_{\nu}U_{\alpha}{}^{[\mu\nu]}=2\sqrt{-g}\,G_{\alpha}{}^{\mu},\quad{}
\partial_{\nu}U_{\alpha}{}^{[\mu\nu]}=2\sqrt{-g}\kappa\,t_{\alpha}{}^{\mu}.\label{9bMar2015}
\end{eqnarray}
The first part indicates inside matter and the second piece
belongs to vacuum gravity~\cite{MTW}. For the condition of
interior mass-energy density, it is known all the classical
pseudotensors give the standard result $2G_{\alpha}{}^{\mu}$, but
only M$\o$ller failed~\cite{CQGSoNesterChen2009}, i.e.,
$\partial_{\nu}U_{\alpha}{}^{\mu\nu}=R_{\alpha}{}^{\mu}$. More
precisely, though the M$\o$ller pseudotensor gives the desired
tidal work, it is disqualified as a satisfactory description of
energy-momentum. Although there are infinite numbers of
pseudotensors, we can remove some forbidden
freedom~\cite{CQGSo2009,PRDSoNester2009}. Thus, perhaps, the Freud
superpotential~\cite{Einstein} is a simpler expression to
demonstrate the tidal work:
\begin{eqnarray}
_{F}U_{\alpha}{}^{[\mu\nu]}=\sqrt{-g}[(\delta^{\mu}_{\alpha}\Gamma^{\lambda\nu}{}_{\lambda}
+\delta^{\nu}_{\alpha}\Gamma^{\mu\lambda}{}_{\lambda}
+\Gamma^{\nu\mu}{}_{\alpha})-(\mu\leftrightarrow\nu)].\label{16aApril2015}
\end{eqnarray}

There are three kinds of superpotential:
$U_{\alpha}{}^{[\mu\nu]}$, $U^{\alpha[\mu\nu]}$ and
$H^{[\alpha\beta][\mu\nu]}$. The total energy-momentum complex can
be obtained as follows
\begin{eqnarray}
2\kappa{\cal{T}}_{\alpha}{}^{\mu}=\partial_{\nu}U_{\alpha}{}^{[\mu\nu]},\quad{}
2\kappa{\cal{T}}^{\alpha\mu}=\partial_{\nu}U^{\alpha[\mu\nu]},\quad{}
2\kappa{\cal{T}}^{\alpha\mu}=\partial^{2}_{\beta\nu}H^{[\alpha\beta][\mu\nu]}.
\end{eqnarray}
In vacuum, $t_{0}{}^{j}$ is the gravitational energy flux and the
tidal work can be computed in either ways
\begin{eqnarray}
\frac{dW}{dt}=\frac{1}{2\kappa}\oint_{\partial{}V}\sqrt{-g}\,t_{0}{}^{j}\,\hat{n}_{j}\,r^{2}\,d\Omega,
\quad
\frac{dW}{dt}=-\frac{1}{2\kappa}\oint_{\partial{}V}\sqrt{-g}\,t^{0j}\,\hat{n}_{j}\,r^{2}\,d\Omega,
\end{eqnarray}
where $r\equiv\sqrt{\delta_{ab}x^{a}x^{b}}$ is the distance from
the body in its local asymptotic rest frame and
$\hat{n}_{j}\equiv{}x_{j}/r$ is the unit radial vector. In our
tidal heating calculation, we will use the deDonder (harmonic)
gauge~\cite{Purdue},
$\partial_{\beta}(\sqrt{-g}g^{\alpha\beta})=0$, which is
equivalently represented as $\Gamma^{\alpha\beta}{}_{\beta}=0$.
From this, we deduce the following identity when $\alpha=0$:
\begin{eqnarray}
\frac{\partial{}h_{00}}{\partial{}x^{0}}=\frac{1}{2}\eta^{cd}\frac{\partial{}h_{0c}}{\partial{}x^{d}}.\label{27dMar2015}
\end{eqnarray}
The metric tensor can be decomposed as
$g_{\alpha\beta}=\eta_{\alpha\beta}+h_{\alpha\beta}$, and its
inverse $g^{\alpha\beta}=\eta^{\alpha\beta}-h^{\alpha\beta}$. Then
we have the following physical expressions~\cite{Purdue}:
\begin{eqnarray}
h_{00}=\frac{2M}{r}+\frac{3}{r^{5}}I_{ij}x^{i}x^{j}-E_{ij}x^{i}x^{j},\quad{}
h_{0j}=\frac{2}{r^{3}}\dot{I}_{ij}x^{i}-\frac{10}{21}\dot{E}_{ik}x^{i}x^{k}x_{j}+\frac{4}{21}\dot{E}_{ij}x^{i}r^{2},
\label{27cMar2015}
\end{eqnarray}
note that $h_{ij}=\delta_{ij}h_{00}$.  Here we remark that
(\ref{27dMar2015}) can be obtained directly from differentiation
using (\ref{27cMar2015}).

\section{Papapetrou, Weinberg, Goldberg and general expression pseudotensors}
Here we consider the Papapetrou, Weinberg and Goldberg
pseudotensors, we also constructed a general pseudotensor
expression. We find that all of them give the same desired tidal
work rate.

\subsection{Papapetrou pseudotensor}
The Papapertrou superpotential is defined as
$H^{[\mu\nu][\alpha\beta]}_{P}
:=-\sqrt{-g}g^{\rho\pi}\eta^{\tau\gamma}\delta_{\pi\gamma}^{\nu\mu}\delta_{\rho\tau}^{\alpha\beta}$,
equivalently one can use
$U^{\alpha[\mu\nu]}_{P}\equiv\partial_{\beta}H^{[\mu\nu][\alpha\beta]}$.
In terms of the Bergmann-Thomson superpotential
$U_{BT}^{\alpha[\mu\nu]}$, we have
\begin{eqnarray}
U_{P}^{\alpha[\mu\nu]}:=U_{BT}^{\alpha[\mu\nu]}-\sqrt{-g}(g^{\rho\tau}h^{\pi\gamma}\Gamma^{\sigma}{}_{\lambda\pi}
+g^{\rho\pi}h^{\sigma\tau}\Gamma^{\gamma}{}_{\lambda\pi})
\delta_{\tau\gamma}^{\mu\nu}\delta_{\rho\sigma}^{\lambda\alpha},
\end{eqnarray}
where
$U^{\alpha[\mu\nu]}_{BT}:=-\sqrt{-g}g^{\alpha\beta}g^{\pi\sigma}
\Gamma^{\tau}{}_{\lambda\pi}\delta^{\lambda\mu\nu}_{\tau\sigma\beta}$.
The Papapetrou pseudotensor becomes
\begin{eqnarray}
&&t^{\alpha\mu}_{P}\nonumber\\
&=&t^{\alpha\mu}_{BT}+(g^{\alpha\mu}\Gamma^{\lambda}{}_{\lambda\beta}
-2\Gamma^{\alpha\mu}{}_{\beta}-2\Gamma^{\mu\alpha}{}_{\beta})\Gamma^{\beta\nu}{}_{\nu}
+(\Gamma^{\alpha\nu}{}_{\nu}+\Gamma^{\nu\alpha}{}_{\nu})\Gamma^{\mu\pi}{}_{\pi}
\nonumber\\
&&+(g^{\alpha\mu}\Gamma^{\lambda}{}_{\lambda\beta}-\Gamma^{\alpha\mu}{}_{\beta}-\Gamma_{\beta}{}^{\alpha\mu}
-2\Gamma^{\mu\alpha}{}_{\beta})\Gamma^{\nu\beta}{}_{\nu}
+(\Gamma^{\alpha\beta\nu}+\Gamma^{\beta\nu\alpha})\Gamma^{\mu}{}_{\beta\nu}
+(\Gamma^{\alpha\beta\nu}+\Gamma^{\nu\beta\alpha})\Gamma_{\beta\nu}{}^{\mu}\nonumber\\
&&+(g^{\alpha\mu}h^{\lambda\pi}-h^{\alpha\mu}g^{\lambda\pi})\Gamma^{\nu}{}_{\lambda\pi,\nu}
-(g^{\pi\mu}h^{\lambda\nu}-h^{\pi\mu}g^{\lambda\nu})\Gamma^{\alpha}{}_{\lambda\pi,\nu}
-(g^{\alpha\pi}h^{\lambda\nu}-h^{\alpha\pi}g^{\lambda\nu})\Gamma^{\mu}{}_{\lambda\pi,\nu}\nonumber\\
&=&-(\Gamma^{\alpha\mu}{}_{\beta}+\Gamma^{\mu\alpha}{}_{\beta})\Gamma^{\beta\nu}{}_{\nu}
+\Gamma^{\alpha\nu}{}_{\nu}\Gamma^{\pi\mu}{}_{\pi}
+\Gamma^{\nu\alpha}{}_{\nu}\Gamma^{\mu\pi}{}_{\pi}
+g^{\alpha\mu}(\Gamma^{\nu\pi}{}_{\rho}\Gamma^{\rho}{}_{\pi\nu}
+\Gamma^{\lambda}{}_{\lambda\beta}\Gamma^{\nu\beta}{}_{\nu})\nonumber\\
&&-2(\Gamma^{\alpha\mu}{}_{\beta}+\Gamma^{\mu\alpha}{}_{\beta})\Gamma^{\nu\beta}{}_{\nu}
+2\Gamma^{\alpha\beta\nu}\Gamma^{\mu}{}_{\beta\nu}
+(g^{\alpha\mu}h^{\lambda\pi}-h^{\alpha\mu}g^{\lambda\pi})\Gamma^{\nu}{}_{\lambda\pi,\nu}\nonumber\\
&&-(g^{\pi\mu}h^{\lambda\nu}-h^{\pi\mu}g^{\lambda\nu})\Gamma^{\alpha}{}_{\lambda\pi,\nu}
-(g^{\alpha\pi}h^{\lambda\nu}-h^{\alpha\pi}g^{\lambda\nu})\Gamma^{\mu}{}_{\lambda\pi,\nu},\label{8aApril2015}
\end{eqnarray}
where the Bergmann-Thomson pseudotensor $t^{\alpha\mu}_{BT}$
expression can be found in~\cite{Favata}. Alternatively, we use
our own representation
\begin{eqnarray}
t^{\alpha\mu}_{BT}
&=&-(g^{\alpha\mu}\Gamma^{\pi}{}_{\pi\rho}-\Gamma^{\mu\alpha}{}_{\rho}
-\Gamma^{\alpha\mu}{}_{\rho})\Gamma^{\rho\,\nu}{}_{\nu}
+\Gamma^{\alpha\nu}{}_{\nu}(\Gamma^{\pi\mu}{}_{\pi}-\Gamma^{\mu\pi}{}_{\pi})
+g^{\alpha\mu}\Gamma^{\nu\rho\pi}\Gamma_{\rho\pi\nu}\nonumber\\
&&+(\Gamma^{\rho\alpha\mu}-\Gamma^{\alpha\mu\rho})\Gamma^{\pi}{}_{\pi\rho}
+\Gamma^{\alpha\pi\rho}(\Gamma^{\mu}{}_{\pi\rho}-\Gamma_{\pi\rho}{}^{\mu})
-\Gamma^{\rho\pi\alpha}(\Gamma^{\mu}{}_{\rho\pi}+\Gamma_{\pi\rho}{}^{\mu}).
\end{eqnarray}
Equation (\ref{8aApril2015}) indicates the gravitational
energy-momentum, for inside matter we have
$t^{\alpha\mu}_{P}=2G^{\alpha\mu}$. The Papapetrou pseudotensor
is, in some sense, a modification of Bergmann-Thomson pseudotensor
by introducing a flat metric $\eta_{\alpha\beta}$. The
corresponding tidal work for Papapetrou pseudotensor is
\begin{eqnarray}
\frac{dW_{P}}{dt}=-\frac{1}{10}\frac{d}{dt}({I}_{ij}{E}^{ij})-\frac{1}{2}\dot{I}_{ij}E^{ij}.
\end{eqnarray}
Here we have the desired result for the tidal work
$-\frac{1}{2}\dot{I}_{ij}E^{ij}$ and the interaction energy
$E_{\rm{}int}=\frac{\gamma+2}{10}I_{ij}E^{ij}$ requires
$\gamma=-3$. Even though the Papapetrou and Landau-Lifshitz
pseudotensors have different expressions, they have the same
choice for the energy localization. Moreover, like the
Landau-Lifshitz pseudotensor, the Papapetrou pseudotensor also a
symmetric pseudotensor.  This implies that the construction of a
conserved total energy-momentum complex is allowed~\cite{Favata}.

\subsection{Weinberg pseudotensor}
The superpotential for Weinberg(W) is
$H^{[\mu\nu][\alpha\beta]}_{W}
:=-\sqrt{-\eta}(\eta^{\xi\rho}\eta^{\kappa\tau}-\frac{1}{2}\eta^{\xi\kappa}\eta^{\rho\tau})
g_{\rho\tau}\eta^{\lambda\sigma}\delta^{\mu\nu}_{\xi\lambda}\delta^{\alpha\beta}_{\kappa\sigma}$.
Alternative representation for this superpotential is
\begin{eqnarray}
U^{\alpha[\mu\nu]}_{W}:=U^{\alpha[\mu\nu]}_{W_{0}}
\quad\quad\quad\quad\quad\quad\quad\quad\quad\quad\quad\quad\quad\quad\quad\quad\quad\quad\quad\quad
\quad\quad\quad\quad\quad\quad\quad\quad
\nonumber\\
+\left\{\sqrt{-\eta}\left[
\begin{array}{cccc}
(g^{\alpha\mu}h^{\nu\,\beta}+h^{\alpha\mu}g^{\nu\beta})\Gamma^{\lambda}{}_{\beta\lambda}
+g^{\alpha\mu}h^{\rho\,\tau}\,\Gamma_{\rho\tau}{}^{\nu}
+g^{\alpha\nu}h^{\mu\rho}\,\Gamma_{\rho}{}^{\beta}{}_{\beta}\\
+(g^{\alpha\nu}h^{\beta\lambda}+h^{\alpha\nu}g^{\beta\lambda}
-g^{\alpha\lambda}h^{\nu\,\beta}-h^{\alpha\lambda}g^{\nu\beta})\Gamma^{\mu}{}_{\beta\lambda}
+h^{\nu\rho}\,\Gamma_{\rho}{}^{\alpha\mu}\\
\end{array}
\right]-(\mu\leftrightarrow\nu)\right\},
\end{eqnarray}
where $U^{\alpha[\mu\nu]}_{W_{0}}
:=-\sqrt{-\eta}\,g^{\alpha\beta}g^{\pi\sigma}
\Gamma^{\tau}{}_{\lambda\pi}\delta^{\lambda\mu\nu}_{\tau\sigma\beta}$.
The Weinberg pseudotensor is
\begin{eqnarray}
t^{\alpha\mu}_{W}=(2\Gamma^{\beta\alpha\mu}-g^{\alpha\mu}\Gamma^{\beta\pi}{}_{\pi})\Gamma_{\beta}{}^{\nu}{}_{\nu}
+g^{\alpha\mu}\Gamma^{\rho\beta\nu}\Gamma_{\rho\beta\nu}
-2\Gamma^{\beta\nu\alpha}\,\Gamma_{\beta\nu}{}^{\mu}
+2h^{\beta\nu}R^{\alpha}{}_{\beta}{}^{\mu}{}_{\nu},\label{9aMar2015}
\end{eqnarray}
which is symmetric in $\alpha$ and $\mu$.  Inside matter we have
$t^{\alpha\mu}_{W}=2G^{\alpha\mu}$. In vacuum we have the tidal
work rate
\begin{eqnarray}
\frac{dW_{W}}{dt}=\frac{1}{10}\frac{d}{dt}(I_{ij}{E}^{ij})-\frac{1}{2}\dot{I}_{ij}E^{ij}.
\end{eqnarray}
Again we have the desired tidal work
$-\frac{1}{2}\dot{I}_{ij}E^{ij}$ and the interaction energy
$E_{\rm{}int}=\frac{\gamma+2}{10}I_{ij}E^{ij}$ indicates
$\gamma=-1$.

\subsection{Goldberg pseudotensor}
Goldberg had changed the weighting factor $\sqrt{-g}$ to be an
arbitrary class density for Einstein and Landau-Lifshitz
pseudotensors~\cite{Goldberg,CQGSoNesterChen2009}.  We discovered
that the tidal work remains unchange if changing the weighting
factor density $(\sqrt{-g}\,)^{n}$, where $n$ is real and finite.
However, it does change the sign of internal energy and
interaction energy. The detail calculation are follows: the first
case of the Goldberg superpotential:
\begin{eqnarray}
_{G}U_{\alpha}{}^{[\mu\nu]}:=-\sqrt{-g}\,^{n}g^{\beta\sigma}\Gamma^{\tau}{}_{\lambda\beta}
\delta^{\lambda\mu\nu}_{\tau\sigma\alpha},
\end{eqnarray}
The corresponding gravitational energy-momentum density is
\begin{eqnarray}
_{G}t_{\alpha}{}^{\mu}=\sqrt{-g}\,^{n}\left[
\begin{array}{cccc}
n(\Gamma^{\beta}{}_{\beta\alpha}\Gamma^{\mu\lambda}{}_{\lambda}
-\delta^{\mu}_{\alpha}\Gamma^{\beta}{}_{\beta\nu}\Gamma^{\nu\lambda}{}_{\lambda}
+\Gamma^{\nu\mu}{}_{\alpha}\Gamma^{\beta}{}_{\beta\nu}
-\Gamma^{\beta}{}_{\beta\alpha}\Gamma^{\lambda\mu}{}_{\lambda})
-2\Gamma^{\beta\nu}{}_{\alpha}\Gamma^{\mu}{}_{\beta\nu}\\
+\delta^{\mu}_{\alpha}\Gamma^{\beta\lambda}{}_{\nu}\Gamma^{\nu}{}_{\beta\lambda}
+(n-1)\delta^{\mu}_{\alpha}\Gamma^{\nu}{}_{\nu\beta}\Gamma^{\lambda\beta}{}_{\lambda}
-(n-2)\Gamma^{\mu}{}_{\alpha\beta}\Gamma^{\nu\beta}{}_{\nu}\quad\quad\quad\quad~\\
\end{array} \right]
\end{eqnarray}
Checking the value of inside matter, we have the standard result
$_{G}t_{\alpha}{}^{\mu}=2G_{\alpha}{}^{\mu}$. The corresponding
tidal work is
\begin{eqnarray}
\frac{dW_{G}}{dt}=\frac{5-2n}{10}\frac{d}{dt}(I_{ij}E^{ij})-\frac{1}{2}\dot{I}_{ij}E^{ij}.\label{18aMay2015}
\end{eqnarray}
Clearly, equation (\ref{18aMay2015}) recovers the result for the
Einstein pseudotensor when $n=1$.

Similarly, the second case of the Goldberg superpotential is
\begin{eqnarray}
U^{\alpha[\mu\nu]}_{G}
:=-(-g)^{n}\,g^{\alpha\beta}g^{\pi\sigma}\Gamma^{\tau}{}_{\lambda\pi}\delta^{\lambda\mu\nu}_{\tau\sigma\beta}.
\end{eqnarray}
The corresponding gravitational energy-momentum density is
\begin{eqnarray}
&&t^{\alpha\mu}_{G}\nonumber\\
&=&(-g)^{n}\left\{
(\Gamma^{\alpha\mu}{}_{\beta}+\Gamma^{\mu\alpha}{}_{\beta}
-2ng^{\alpha\mu}\Gamma^{\gamma}{}_{\gamma\beta})\Gamma^{\beta\nu}{}_{\nu}
+[(2n-1)\Gamma^{\nu\alpha}{}_{\nu}-\Gamma^{\alpha\nu}{}_{\nu}]\Gamma^{\mu\pi}{}_{\pi}
+\Gamma^{\alpha\nu}{}_{\nu}\Gamma^{\pi\mu}{}_{\pi}
\right\}\nonumber\\
 &&+(-g)^{n}\left\{
\begin{array}{cccc}
[(2n-1)(g^{\alpha\mu}\Gamma^{\gamma\nu}{}_{\gamma}-\Gamma^{\mu\alpha\nu})
+2n\Gamma^{\nu\alpha\mu}-\Gamma^{\alpha\mu\nu}]\Gamma^{\pi}{}_{\nu\pi}
+g^{\alpha\mu}\Gamma^{\beta\nu}{}_{\rho}\Gamma^{\rho}{}_{\beta\nu}\\
+(1-2n)\Gamma^{\nu\alpha}{}_{\nu}\Gamma^{\pi\mu}{}_{\pi}
+(\Gamma^{\alpha\beta}{}_{\nu}-\Gamma^{\beta\alpha}{}_{\nu})\Gamma^{\mu\nu}{}_{\beta}
-(\Gamma^{\alpha\nu}{}_{\beta}+\Gamma^{\nu\alpha}{}_{\beta})\Gamma^{\beta\mu}{}_{\nu}\\
\end{array}
\right\}.
\end{eqnarray}
Inside matter we have $t^{\alpha\mu}_{G}=2G^{\alpha\mu}$ and the
tidal work in vacuum is
\begin{eqnarray}
\frac{dW_{G}}{dt}=\frac{3-4n}{10}\frac{d}{dt}(I_{ij}E^{ij})-\frac{1}{2}\dot{I}_{ij}E^{ij}.
\end{eqnarray}
Once again, we recover the result for the Landau-Lifshitz
pseudotensor when $n=1$.

\subsection{General pseudotensor expression}
Thorne had laid down a statement that all pseudotensors give the
same tidal work rate~\cite{Thorne}. This means that the tidal work
expression is pseudotensor independent of one's choice how to
localize the gravitational energy. As far as the classical
pseudotensors are concerned, after the examinations by
Purdue~\cite{Purdue}, Favata~\cite{Favata} and the present paper,
we clarified that Thorne's prediction is correct. Here we
demonstrate that it is still true even for a general form of
pseudotensor. For simplicity, we use Freud superpotential as a
leading term and consider the 13 linear artificial higher order
terms combination as a modification:
\begin{eqnarray}
U_{\alpha}{}^{[\mu\nu]}:={}_{F}U_{\alpha}{}^{[\mu\nu]}+\sqrt{-g}\sum_{i=1}^{13}k_{i}U_{i},\label{23Mar2016}
\end{eqnarray}
where $k_{i}$ are finite constants and the explicit extra terms
are
\begin{eqnarray}
&&U_{1}=\delta^{\mu}_{\alpha}h^{\beta\nu}\Gamma^{\lambda}{}_{\lambda\beta},\quad{}
U_{2}=h^{\mu\pi}\Gamma_{\alpha\pi}{}^{\nu},\quad\quad{}
U_{3}=\delta^{\mu}_{\alpha}h^{\beta\lambda}\Gamma^{\nu}{}_{\beta\lambda},\quad{}
U_{4}=h_{\alpha}{}^{\mu}\Gamma^{\lambda\nu}{}_{\lambda},\quad{}\nonumber\\
&&U_{5}=h_{\alpha\lambda}\Gamma^{\nu\mu\lambda},\quad\quad{}
U_{6}=\delta^{\mu}_{\alpha}h\Gamma^{\lambda\nu}{}_{\lambda},\quad\quad{}
U_{7}=\delta^{\mu}_{\alpha}h^{\beta\lambda}\Gamma_{\beta\lambda}{}^{\nu},\quad{}
U_{8}=h\Gamma^{\nu\mu}{}_{\alpha},\nonumber\\
&&U_{9}=\delta^{\mu}_{\alpha}h\Gamma^{\nu\lambda}{}_{\lambda},\quad~~{}
U_{10}=\delta^{\mu}_{\alpha}h^{\beta\nu}\Gamma_{\beta}{}^{\lambda}{}_{\lambda},~~{}
U_{11}=h_{\alpha}{}^{\mu}\Gamma^{\nu\lambda}{}_{\lambda},\quad~~{}
U_{12}=h^{\mu\pi}\Gamma^{\nu}{}_{\alpha\pi},\nonumber\\
&&U_{13}=h^{\mu\pi}\Gamma_{\pi\alpha}{}^{\nu}.
\end{eqnarray}
where $\mu$ and $\nu$ are anti-symmetric such that every $U_{i}$
is subtracted by an extra term follows the swaped $\mu$ and $\nu$
indices. In (\ref{23Mar2016}), consider inside matter, the leading
term Freud superpotential gives the standard value
$2G_{\alpha}{}^{\mu}$, but all $U_{i}$ contribute null result.
However, we find $U_{1}$ to $U_{8}$ alter the values of the
internal energy and interaction energy, but $U_{9}$ to $U_{13}$
give vanishing values. In other words, $U_{1}$ to $U_{8}$ affect
the choice how to localize the gravitational energy, however
$U_{9}$ to $U_{13}$ do not. One minor thing might need to clarify
that if we change the density of the weighting factor $\sqrt{-g}$
in (\ref{23Mar2016}), it does not affect the result of the tidal
work as mentioned in section 3.3. After a straightforward
manipulation, the tidal work is
\begin{eqnarray}
\frac{dW}{dt}=\frac{3-\alpha}{10}\frac{d}{dt}(I_{ij}E^{ij})-\frac{1}{2}\dot{I}_{ij}E^{ij},\label{27Mar2015}
\end{eqnarray}
provided
$\alpha=k_{1}+k_{2}-k_{3}-k_{4}-k_{5}+2k_{6}+2k_{7}+2k_{8}$. Thus,
equation (\ref{27Mar2015}) illustrates that the gravitational
energy localization is $E_{\rm{}int}$ dependence, while the tidal
work rate is unique.  Thorne's prediction is correct.

\section{Conclusion}
Thorne claimed that all pseudotensors give the same tidal work as
the Newtonian theory. Purdue used the Landau-Lifshitz classical
pseudotensor to calculate the tidal work and the result matched
with the Newtonian perspective. Later Favata employed the same
method to examine the Einstein, Bergmann-Thomson and M{\o}ller
classical pseudotensors, all of them give the same result as
Purdue did. Inspired by the work of Purdue and Favata, for the
completeness, we take into account the rest of the three classical
pseudotensors: Papapetrou, Weinberg and Goldberg.  We find that
these three pseudotensors give the same desired result as Purdue
obtained. In addition, we emphasize that a suitable gravitational
energy-momentum pseudotensor requires satisfy the inside matter
condition and all of them pass this test but only M$\o$ller
failed. Moreover, we constructed a general pseudotesnor expression
which is modified by 13 linear artificial higher order terms
combination with the Einstein pseudotensor. This general
pseudotensor fulfills the conditions of inside matter.
Consequently, we have verified that Thorne's assertion is valid,
i.e., tidal work is independent of one's choice how to localize
the gravitational energy for all kinds of pseudotensor.


\end{document}